# A Hierarchical Framework for Ambient Signals based Load Modeling with Exploring the Hidden Quasi-convexity

Xinran Zhang, *Member, IEEE*, David J. Hill, *Life Fellow, IEEE*, Chao Lu, *Senior Member, IEEE*, and Yue Song, *Member, IEEE*

*Abstract*—Load modeling is an important issue in modeling a power system. The approach of ambient signals-based load modeling (ASLM) was recently proposed to better track the time-varying changes of load models. To improve computation efficiency and model structure complexity, a hierarchical framework for ASLM is proposed in this paper. Through this framework, the hidden quasi-convexity of load modeling problem is explored for the first time, and more complicated static load model structures can be applied. In the upper stage, the identification of dynamic load parameters is regarded as an optimization problem. In the lower stage, the optimal static load parameters are obtained through linear regression for a given group of dynamic load parameters. Afterwards, the regression residuals are regarded as the objective function (OF) of the upper stage optimization problem. The proposed method is validated by the case study results in Guangdong Power Grid. The results have shown that the OF is mostly quasi-convex after the transformation of induction motor model, which provides the basis for the application of gradient-based optimization algorithm. The case study results also validate that the proposed approach has better computation efficiency and model structure complexity compared with the previous ASLM approach.

*Index Terms*--Load modeling, parameter identification, ambient signals, convex optimization.

## I. INTRODUCTION

LOAD modeling has been an important and challenging issue in power system [1]. The accuracy of power system time-domain simulation relies on the accuracy of power system models, which means inappropriate power system models may lead to misleading simulation results [2]. Unlike most of the other power system models, load model is an aggregation of multiple different types of load components, which makes load modeling more complicated [3].

Two categories of load modeling methodologies have been proposed in previous research, i.e., the component based approach and the measurement based approach [4]. Being more widely used, the measurement based load modeling is to consider all the components connected to a load bus as a composite load model. Then, the parameters are identified from local measurement data within a pre-selected model structure.

This paper was supported by a grant from the Research Grants Council of the Hong Kong Special Administrative Region under Theme-based Research Scheme through Project No. T23-701/14-N.

X. Zhang, D. J. Hill and Y. Song are with the Department of Electrical and Electronic Engineering, The University of Hong Kong, Hong Kong (e-mail: zhangxr@eee.hku.hk, dhill@eee.hku.hk, songyue@eee.hku.hk). C. Lu is with the State Key Laboratory of Control and Simulation of Power System and Generation Equipment, Tsinghua University, Beijing 100084 (e-mail: luchao@tsinghua.edu.cn ).

In the previous measurement based load modeling research, the composite load model with the ZIP model as the static load part and the induction motor (IM) as the dynamic load part is widely used, forming the well-known ZIP+IM model [5]. In recent years, with the increasing penetration of renewable energy in the demand side of power systems, the distributed renewable energy components and power electronics loads are considered in load modeling in [6], [7]. The robust time-varying load modeling approach has been proposed in [8], [9] to continuously track the load model parameters with the previously estimated parameters as the initial values. In addition, more complicated load model structure, such as the Western Electricity Coordinating Council load model, has been applied in the research of load modeling [10].

In the previous measurement based load modeling approaches, the mostly used measurement data source is the post large disturbance response (PLDR) data. Therefore, when the PLDR based load modeling can be conducted depends on the occurrence of large disturbance events. Then, PLDR based load modeling cannot be frequently conducted because large disturbance events only occasionally happen in power systems. Recently, with the increasing variety of load components and the increasing integration of uncertain power resources such as renewables and demand response, it is more necessary to track the time-varying property of load models, which cannot be achieved by the PLDR based load modeling.

Recently, ambient signals based load modeling (ASLM) approach is proposed to better track the time-varying property of load models [11], [12]. Ambient signals refer to the small disturbances contained in power and voltage signals during power system daily operation [13]. With ambient signals always existing in power system measurements, ASLM can be conducted almost at any time, without the dependence on the occurrence of large disturbance events. In this way, load models can be identified periodically to better track the time-varying changes. A common weakness in previous work of ASLM is that the static load model structure is simplified as constant Z to reduce the computation complexity, which should be improved for more generalized and more commonly used static load model structures.

Optimization algorithms have been widely used to identify the load model parameters. Another weakness of the previous measurement load modeling approaches is the lack of enough consideration about the convexity of the objective functions (OFs) in the optimization problems. In [11], the OF of the induction motor model has been simulated to be non-convex.

As a result, the heuristic optimization algorithms, which have stronger global optimal solution (GOS) searching capability in non-convex problems, have been widely applied previously, including genetic algorithm [5], support vector machines [4], simulated annealing [2], and differential evolution [6], [11]. Nevertheless, if the optimization problem can be approximately convexified, the computation efficiency and robustness can be significantly improved by adopting gradient-based algorithms.

The contributions of this paper are summarized as follows. Firstly, a hierarchical framework for ASLM is proposed. The dynamic load model parameters are identified in the upper stage through gradient-based optimization, and the static load model parameters are identified in the lower stage through regression. Secondly, the OF in the upper stage optimization problem is quasi-convexified in most parts of the feasible region through the transformation of the induction motor model, which provides the theoretical basis for the application of gradient-based algorithm. This is the first time that load modeling is solved as a quasi-convex optimization problem with gradient-based algorithm. As a result, the computation efficiency is significantly improved. Thirdly, compared with the previous ASLM approaches, the structure of the static load model is improved from the constant Z model to the more widely used ZIP model, and still has the potential to apply other static load model structures such as the exponential model. Through the improvement of computation efficiency and model structure complexity, the practical applicability and soundness of the recently proposed ASLM approach are significantly enhanced.

The rest of this paper is organized as follows. The composite load model structure and the transformation of the induction motor model are discussed in Section II. In Section III, the hierarchical framework of ASLM is proposed. In Section IV, the regression method to identify the static load model in the lower stage is proposed. In section V, the gradient-based algorithm to solve the optimization problem in the upper stage in order to identify the dynamic load model parameters is proposed. In Section VI, the case study results are presented to validate the proposed hierarchical framework and the quasi-convexity of the problem. Section VII concludes this paper.

## II. LOAD MODEL STRUCTURE

### A. Composite Load Model Structure

For a load bus in a power system, there are four local measurement signals which are used in load modeling, i.e., the bus voltage magnitude ($V$) and phase angle ($\theta$), the active power ($P$) and the reactive power ($Q$). The function of a load model is to predict $P$ and $Q$ under a given group of voltage phasors ($V$ and $\theta$). Generally, there are two parts in a composite load model structure, i.e., the static load part and the dynamic load part. The relationship between $P$, $Q$ and $V$ of static load models is described by algebraic equations. In contrast, the relationship between $P$, $Q$ and $V$, $\theta$ of dynamic load models is described by differential equations and algebraic equations.

### B. Dynamic Load Model: Induction Motor

#### 1) Third-order Model

The parameters of the third-order induction motor model are introduced as follows: $X$ is the rotor open circuit reactance, $X'$ is the rotor transient reactance, $T_{d0}$ is the rotor open-circuit time constant, $H_2$ is the inertia time constant, $T_m$ is the mechanical torque, which is assumed to be constant in this paper. Then, there are five parameters in this model, i.e. [$X$ $X'$ $T_{d0}$ $H_2$ $T_m$]. The third-order state-space formulae of the induction motor are given as follows,

$$\begin{cases} \dfrac{dE_d}{dt} = -\dfrac{X}{T_{d0}X'}E_d + s\omega_0 E_q + \dfrac{1}{T_{d0}}(\dfrac{X}{X'}-1)V_d \\ \dfrac{dE_q}{dt} = -\dfrac{X}{T_{d0}X'}E_q - s\omega_0 E_d + \dfrac{1}{T_{d0}}(\dfrac{X}{X'}-1)V_q \\ \dfrac{ds}{dt} = \dfrac{1}{H_2}(T_m - T_e) \end{cases} \quad (1)$$

where $E_d$ and $E_q$ are the d-axis and q-axis components of the electromotive force phasor, $V_d$ and $V_q$ are the d-axis and q-axis components of the voltage phasor, $\omega_0$ is the synchronous rotation angular speed, $T_e$ is the electromagnetic torque, which is calculated as follows,

$$T_e = \dfrac{E_d V_q - E_q V_d}{X'} \quad (2)$$

Then, $P$ and $Q$ can be calculated from $E_d$, $E_q$, $V_d$ and $V_q$ as follows, which form the output formulae of the induction motor:

$$\begin{aligned} P &= \dfrac{E_d V_q - E_q V_d}{X'} \\ Q &= \dfrac{V_d^2 + V_q^2}{X'} + \dfrac{-V_d E_d - V_q E_q}{X'} \end{aligned} \quad (3)$$

#### 2) Model Transformation

The form of the state-space formulae in (1) is complicated, which is part of the reason for the complexity and difficulty of load modeling problem. In order to simplify the form of (1) and reduce the number of parameters, the third-order model is transformed. Firstly, two new state variables are defined, i.e., $F_d = E_d/X'$ and $F_q = E_q/X'$, to replace the original state variables $E_d$ and $E_q$. By defining the new parameters $a = \dfrac{1}{T_{d0}X'}(\dfrac{X}{X'}-1)$ and $b = \dfrac{X}{T_{d0}X'}$, the state space formulae are transformed as follows,

$$\begin{cases} \dfrac{dF_d}{dt} = -bF_d + s\omega_0 F_q + aV_d \\ \dfrac{dF_q}{dt} = -bF_q - s\omega_0 F_d + aV_q \\ \dfrac{ds}{dt} = \dfrac{1}{H_2}(T_m - V_q F_d + V_d F_q) \end{cases} \quad (4)$$

It can be observed that the state space formulae of the induction motor model have been simplified compared with the original form. Then, the next task is to calculate $P$ and $Q$ from the newly defined state variables $F_d$ and $F_q$. From (3), it can be observed that the first part of $Q$, $(V_d^2+V_q^2)/X'$, has the same form with the static load models, because it is equal to $V^2/X'$ and is only related to $V$. If this part of $Q$ is regarded as a part of the static load, the rest part of $Q$, together with $P$, can be calculated from $V_d$, $V_q$, $F_d$ and $F_q$ as follows,

$$\begin{aligned} P &= F_d V_q - F_q V_d \\ Q &= -V_d F_d - V_q F_q \end{aligned} \quad (5)$$

In this way, the number of parameters is reduced. There are four parameters to be identified in the transformed induction motor model, i.e. [$a$ $b$ $H_2$ $T_m$]. The initial motivation of transforming the induction motor model is to reduce the number of parameters and simplify the state-space formulae. However, it will be validated in the Section VI that the transformation has contributed to improving the convexity of the load modeling problem.

*C. Static Load Model: ZIP*

In this paper, the widely used ZIP model is chosen as the static load model. The relationship between *P*, *Q* and *V* of the ZIP model is described as follows,

$$\begin{cases} P = P_z V^2 + P_i V + P_p \\ Q = Q_z V^2 + Q_i V + Q_p \end{cases} \quad (6)$$

where there are altogether six parameters to be identified, i.e. [$P_z$, $P_i$, $P_p$, $Q_z$, $Q_i$, $Q_p$]. Here $Q_z$ has included $1/X'$ caused by the transformation of the induction motor model. Apart from the ZIP model, other static load models whose parameters can be identified through regression can also be applied in this structure, such as the exponential load model.

## III. AMBIENT SIGNALS BASED LOAD MODELING: CONCEPT AND HIERARCHICAL FRAMEWORK

*A. Ambient Signals based Load Modeling*

As a branch of measurement based load modeling, ASLM is to identify load model parameters from the measurement data of ambient signals. The spectrum of ambient signals in power systems is distributed between 0.2 and 2.0 Hz, which belongs to the frequency range of power system electromechanical dynamics [13]. Therefore, it is reasonable to apply ambient signals data in the identification of load model parameters.

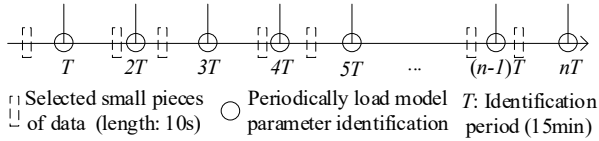

Fig. 1 Timeline of ambient signals-based load modeling

The advantage of the ASLM over the PLDR based load modeling is the achievement of more frequent and periodical tracking of the time-varying changes of load models. A brief timeline of periodical ASLM is given in Fig. 1, where *T* is the identification period, e.g. 15 min. The phasor measurement unit (PMU) measured data is used in identification, the measurement period of which can be as short as 10 ms. At the end of each period, a small amount of data within this period is selected for load modeling, the length of which is much shorter than *T*, e.g. 10 s. The reason for selecting small amounts of data is that load models are assumed not to frequently change significantly within one period. In this way, the time-varying properties of load models can be periodically tracked.

The limitations of ASLM are mainly caused by the relatively smaller disturbance magnitude. As a result, the identification results are more sensitive to PMU measurement errors. In addition, only the partial dynamics around the steady operating point of the load models can be perceived through ambient signals. Nevertheless, the load model identified from ambient signals is practically useful for the following two reasons. Firstly, it is still better to apply the periodically updated load models than to apply the historical identified models from PLDR in order to better track the time-varying changes. Secondly, the models of most load components under large and small disturbance situations are similar, such as induction motors and static load [14]. Therefore, the accuracy of the load models built from ambient signals can be guaranteed.

*B. Hierarchical Identification Framework*

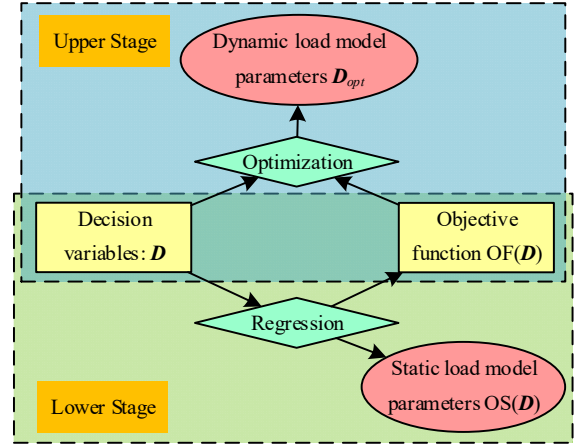

Fig. 2 Hierarchical framework for ambient signals based load modeling

In this paper, a hierarchical framework is proposed to identify load model parameters from ambient signals, as shown in Fig. 2. This framework includes two stages, i.e. an optimization problem in the upper stage, and a regression problem in the lower stage. Through this framework, the parameters of both static and dynamic load models can be identified from the time series measurement data of $V_m$, $\theta_m$, $P_m$ and $Q_m$, where the subscript *m* refers to the measured values.

In the optimization problem in the upper stage, the set of the decision variables ***D*** consists of the induction motor parameters, [*a*, *b*, $H_2$, $T_m$]. The OF is the average of the squared difference between the measured values ($P_m$, $Q_m$) and the model predicted values ($P_p(D)$, $Q_p(D)$) of *P* and *Q*, i.e. (||$P_m$-$P_p(D)$||+||$Q_m$-$Q_p(D)$||)/*l*, where *l* is the length of the data used in identification. However, the parameters of the static load model are not included in ***D***, which are necessary in calculating $P_p$ and $Q_p$.

To solve this problem, a regression problem is solved in the lower stage. With $V_m$, $\theta_m$ and ***D***, the power consumption by the induction motor can be estimated as $P_{p\_im}(D)$ and $Q_{p\_im}(D)$. Then, the rest parts in $P_m$ and $Q_m$ are regarded to be consumed by static load, which are just the dependent variables for the regression problem. Then, a linear regression problem is solved to obtain the optimal static load model parameters for ***D*** (OS(***D***)), after which the OF for ***D*** (OF(***D***)) can be calculated according to the regression residuals.

To conclude, the regression problem in the lower stage is designed to calculate OF(***D***) and OS(***D***) for a given ***D***. Back to the upper stage, with ***D*** being the decision variables and OF(***D***) being the OF, the optimal ***D*** (***D***$_{opt}$) which can minimizes OF(***D***) is regarded as the identification results of dynamic load model parameters. As for the static load model parameters, OS(***D***$_{opt}$) is chosen as the identification results. In this way, all the parameters can be+ identified in this hierarchical framework.

## IV. IDENTIFICATION OF STATIC LOAD MODEL: REGRESSION

In this section, the regression problem in the lower stage to calculate OF(**D**) and OS(**D**) for a given **D** is discussed. There are two sub-problems in this stage, i.e. the prediction of induction motor power consumption, and the regression of static load model parameters, as shown in Fig. 3.

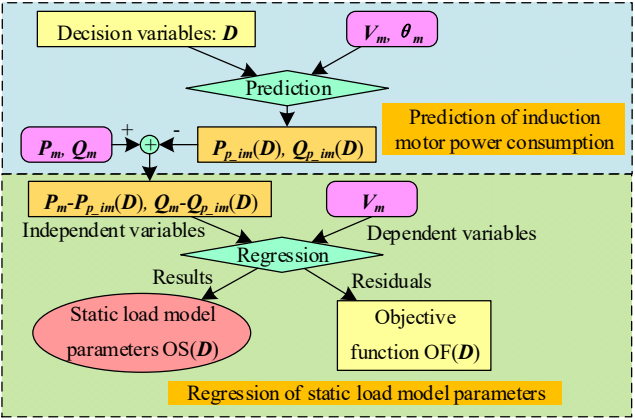

Fig. 3 Flowchart of lower stage regression problem

### A. Prediction of Induction Motor Power Consumption

Firstly, the power consumption of the induction motor $P_{p\_im}$ and $Q_{p\_im}$ should be predicted. In order to predict $P_{p\_im}$ and $Q_{p\_im}$, the following data is required: $V_m$, $\theta_m$, **D**, and the initial states of the induction motor $[F_{d0}, F_{q0}\ s_0]$. $V_m$ and $\theta_m$ can be obtained from PMU measurement data, and **D** is the set of decision variables. However, $[F_{d0}, F_{q0}\ s_0]$ are unknown, the estimation of which is a key problem in this section.

From our simulation experience, we found out a fact that, for two induction motors with the same parameters but different initial states, the dynamics of the state and output variables will be close after about 0.5s. Based on this fact, the initial states $[F_{d0}, F_{q0}\ s_0]$ are estimated by setting the left-hand side of all the equations in (4) as 0, which means the induction motor is assumed to start from a steady state. In addition, the data of the starting period is not used in identification because the estimation of state variables is not accurate enough during this period. In this way, the initial states $[F_{d0}, F_{q0}\ s_0]$ are estimated.

Afterwards, $P_{p\_im}$ and $Q_{p\_im}$ can be predicted from $V_m$, $\theta_m$, **D**, and $[F_{d0}, F_{q0}\ s_0]$ as follows. In the $n$th prediction step, $V_{d,n}$ and $V_{q,n}$ are first computed from $V_{m,n}$ and $\theta_{m,n}$ according to $V_{d,n}+iV_{q,n}=V_{m,n}e^{i\theta m,n}$. Then, $[F_{d,n}, F_{q,n}\ s_n]$ is predicted from $[F_{d,n-1}, F_{q,n-1}\ s_{n-1}]$ according to (4) by the explicit integral method, after which $[P_{p\_im,n}, Q_{p\_im,n}]$ can be calculated according to (5). In this way, $P_{p\_im}(\boldsymbol{D})$ and $Q_{p\_im}(\boldsymbol{D})$ are predicted.

### B. Regression of Static Load Model Parameters

After the prediction of $P_{p\_im}(\boldsymbol{D})$ and $Q_{p\_im}(\boldsymbol{D})$, the rest parts in $P_m$ and $Q_m$ are regarded to be consumed by the static load. Then, OF(**D**) and OS(**D**) are calculated as the functions of a given **D** through regression.

In this paper, the widely used ZIP model is selected as the model structure of the static load. The active power related and reactive power related parameters are identified separately as two regression problems. The active power related parameters are identified through $([P_p, P_i, P_z], r_p)$=regress($P_m-P_{p\_im}(\boldsymbol{D})$, [**1**, $V_m$, $V_m^2$]), where $P_m-P_{p\_im}(\boldsymbol{D})$ is the dependent variable, [**1**, $V_m$, $V_m^2$] are the independent variables, **1** is an all-ones vector with the same size as $V_m$. The static load model parameters are as follows: $P_p$ is the coefficient of **1**, $P_i$ is the coefficient of $V_m$, and $P_z$ is the coefficient of $V_m^2$. In addition, $r_p$ is the series of regression residuals. Similarly, the reactive power related parameters can be obtained through $([Q_p, Q_i, Q_z], r_q)$=regress($Q_m-Q_{p\_im}(\boldsymbol{D})$, [**1**, $V_m$, $V_m^2$]). In this way, OS(**D**) is calculated as $[P_p, P_i, P_z, Q_p, Q_i, Q_z]$.

Another task in solving the regression problem is to calculate OF(**D**). In section III.B, OF(**D**) has been defined, so the essential problem here is the calculation of $(P_p(\boldsymbol{D}), Q_p(\boldsymbol{D}))$. The model predicted power consumption of the induction motor has been predicted as $P_{p\_im}(\boldsymbol{D})$ and $Q_{p\_im}(\boldsymbol{D})$. As for the static load, the model predicted power consumption $P_{p\_st}(\boldsymbol{D})$ and $Q_{p\_st}(\boldsymbol{D})$ are the estimated values of the dependent variables in the regression. Therefore, the residuals in the regression are just the difference between the measured values and the model predicted values of active and reactive power consumption, i.e.: $r_p=P_m-P_{p\_im}(\boldsymbol{D})-P_{p\_st}(\boldsymbol{D})$ and $r_q=Q_m-Q_{p\_im}(\boldsymbol{D})-Q_{p\_st}(\boldsymbol{D})$. Then, OF(**D**) is calculated as follows:

$$OF(\boldsymbol{D}) = (\|r_p\| + \|r_q\|)/l \qquad (7)$$

## V. IDENTIFICATION OF DYNAMIC LOAD MODEL: OPTIMIZATION

### A. Description of Optimization Problem

In this section, the optimization problem is descripted, including the decision variables, the OF and the constraints. The decision variables are the induction motor parameters, **D**=[$a$, $b$, $H_2$, $T_m$]. The OF has been defined as OF(**D**), which is $(\|P_m-P_p(\boldsymbol{D})\|+\|Q_m-Q_p(\boldsymbol{D})\|)/l$.

As for the constraints, two aspects should be included. Firstly, the reasonable ranges should be set for all the parameters in **D**. After testing 350,000 different induction motors, whose parameters are randomly generated according to the ranges of parameters in [5], the ranges of [$a$, $b$, $H_2$] can be approximately given as: [[10, 3, 0.5], [80, 30, 3]]. The range of $T_m$ is within [0, $P_{mean}$] where $P_{mean}$ is the average value of $P_m$. Secondly, to ensure the stable operation of an induction motor according to [15], $a$ and $b$ should obey the following constraint: $aV_{min}^2>2b$, where $V_{min}$ is the minimal value of $V_m$. If this constraint is violated, the maximal $T_e$ of the induction motor will be less than $T_m$, which makes it impossible to maintain the balance between $T_m$ and $T_e$ and the stable operation of the induction motor.

### B. Quasi-convexity

In this section, the concept of quasi-convexity is introduced. The quasi-convexity of a optimization problem is defined as follows: a function $f:S\in\boldsymbol{R}$ defined on a convex subset $S$ of a real vector space is quasi-convex if for all $x, y\in S$ and $\lambda\in[0,1]$, the following inequation holds:

$$f(\lambda x+(1-\lambda)y)<\max\{f(x),f(y)\} \qquad (8)$$

This definition means that it is always true for $f$ that a point directly between two other points does not give a larger function value than both other points do. For a quasi-convex function, if a local minimal solution exists, it is ensured to be the GOS. Based on these properties, gradient-based algorithms can be applied in solving quasi-convex optimization problems.

*C. SQP Optimization Algorithm*

In this paper, the sequential quadratic programming (SQP) algorithm is applied to solve the optimization problem in the upper layer. The procedures of the iteration process can be briefly summarized as follows. Firstly, an initial feasible solution (IFS) $D_0$ is generated within the feasible region according to the constraints. Then, in each iterative step $k$, $k \in N$, the optimization problem is approximately modeled as a quadratic programming (QP) sub-problem in the neighborhood region of the $D_{k-1}$. After the QP subproblem is solved, the solution can be used to construct a new $D_k$. With the increase of $k$, the iterate sequence $(D_k)_{k \in N}$ converges to a local minimum.

## VI. CASE STUDY

*A. Power System Used in Case Studies*

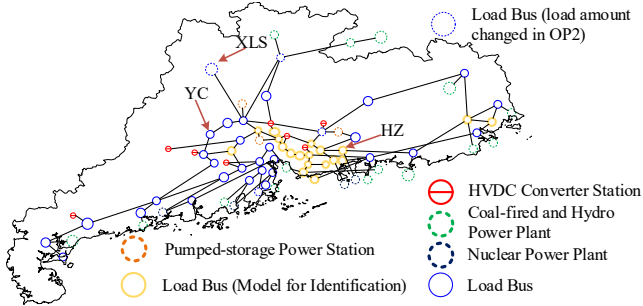

Fig. 4 Structure of Guangdong Power Grid

The 500kV network of the Guangdong Power Grid is used as the test system for load model identification and validation, the structure of which is given in Fig. 4. This system includes 83 buses, 97 lines, 52 load buses and 20 generators. The simulation in this section is conducted in the Power System Analysis Toolbox (PSAT) in MATLAB. The time length of one simulation case (SC) is 10s, with the time step being 0.01 s. The base value of the system capacity is 100 MVA.

The composite load models are applied in all the load buses. The models of 20 load buses are identified in this section. To generate the ambient signals in the system, independent groups of low-pass filtered white noise are penetrated in the active power consumption of 30 other load buses. The cut-off frequency of the low-pass filter is 2 Hz, according to the frequency spectrum of practical measured ambient signals [11].

50 different SCs are conducted to generate the data for load modeling, which are denoted as SC 1 to SC 50. In different SCs, different groups of ambient signals and different load models are applied. Since there are 20 load models to be identified in one SC, there are 1000 load modeling cases (LMC). In one LMC, the parameters of one load model are identified from a group of $V_m$, $\theta_m$, $P_m$ and $Q_m$ measurements.

*B. Identification: Examples*

  *1) An Example without Measurement Errors*

Firstly, an example LMC of how the load model parameters are identified from ambient signals is given. The load model on Bus DG in SC 1 is identified in this LMC. The actual values of the load model parameters are $D_{real}$=[33.98 15.10 1.34 7.33] for [$a$ $b$ $H_2$ $T_m$] and [1.20 1.77 1.59 -2.82 -2.63 -2.36] for [$P_z$ $P_i$ $P_p$ $Q_z$ $Q_i$ $Q_p$]. The measurements of $V_m$, $\theta_m$, $P_m$ and $Q_m$ are given in Fig. 5. The measurement errors are not considered here, which will be discussed in the following sections.

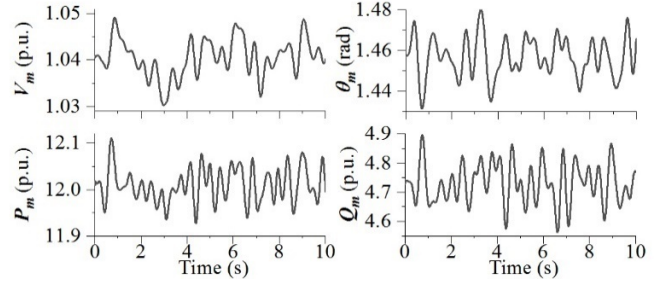

Fig. 5 Measurement curves of Bus DG in SC 1

  *a) Identification*

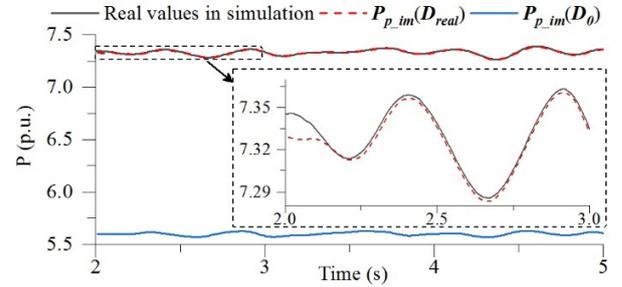

Fig. 6 Prediction results of the P consumption of the induction motor

The method which is proposed in Section IV to identify load model parameters from ambient signals is illustrated in this subsection. The data from 2 s to 10 s is used for identification.

Firstly, the method to predict $P_{p\_im}(D)$ and $Q_{p\_im}(D)$ for a given $D$ is validated. Based on the measurements of $V_m$, $\theta_m$, $P_m$ and $Q_m$ in Fig. 5, $P_{p\_im}(D)$ and $Q_{p\_im}(D)$ are predicted for two different $D$, i.e. $D_{real}$ and $D_0$=[44.50 16.47 0.81 5.60]. The prediction results of $P_{p\_im}(D_{real})$ and $P_{p\_im}(D_0)$ are given in Fig. 6, in which they are compared with the real values of the active power consumption of the induction motor part in simulation. It can be observed that the initial values of $P_{p\_im}(D_{real})$ deviate from the real values because the estimations of the initial states are not absolutely accurate. However, $P_{p\_im}(D_{real})$ will become very close to the real values after a short period, the length of which is shorter than 0.2 s. More similar cases are also tested for other load models, and the conclusion still holds. For comparison, it can be observed that $P_{p\_im}(D_0)$ deviate much more from the real values. Therefore, the method to predict $P_{p\_im}(D)$ and $Q_{p\_im}(D)$ is validated.

Then, the next step is to identify the optimal parameters of the static load model for a given $D$ through regression. In the regression process, the data from 2 s to 3 s is not used to avoid the impact of inaccurate estimation of the initial states. Then, the data between 3 s and 10 s is used in regression to calculate OF($D$) and OS($D$). The regression for $D_0$ and $D_{real}$ is given as examples here. For $D_{real}$, OS($D_{real}$)=[3.04 -2.06 3.58 23.36 -5.40 -0.92], and OF($D_{real}$)=1.36*10$^{-6}$. For $D_0$, OS($D_0$)=[-34.48 78.70 -38.11 -141.89 337.21 -178.37], and OF($D_0$)=1.16*10$^{-3}$. Smaller OF($D$) can be obtained with $D$ approaching the actual values of the parameters, which means OF($D$) is suitable to be used as the OF in the optimization of **D**.

Finally, the parameters of the induction motor are identified through optimization. The IFS is selected to be $D_0$. After 21 iterations in the SQP, the convergence condition of the total changes of the decision variables is met, after which $D_{opt}$=[33.91  15.07  1.33  7.33], OF($D_{opt}$)=1.35*10$^{-6}$, and OS($D_{opt}$)=[2.18 -0.24 2.62 22.61 -3.83 -1.75]. $D_{opt}$ is very close to $D_{real}$, which means the parameters of the induction motor part are accurately identified. As for the static load parameters, the identified $Q_z$ has included $X'$, which makes it different from the real values. Although OS($D_{opt}$) is not the same as the real static load parameters, the total amount of static load can be ensured because $T_m$ is accurately identified.

*b) Validation*

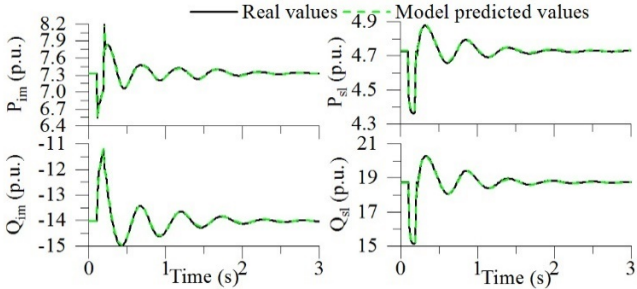

Fig. 7 Validation results of Bus DG in SC1. $P_{im}$, $Q_{im}$: power consumption of induction motor. $P_{sl}$, $Q_{sl}$: power consumption of static load.

Following the previous subsection, whether the identified load models can represent the dynamics of the original load models under large disturbance events should be validated, since the static load model parameters are not the same. In this section, two three-phase to ground faults on Bus YC and Bus XLS are selected as the validation events. The operating point (OP) is changed from the initial OP1 to a new OP OP2, in which the load amounts of four load buses are doubled, as shown in Fig. 4. Two scenarios are simulated for one validation case. In Scenario 1, the actual load models are used. In Scenario 2, the models of the 20 load buses are replaced with the identified models. The similarity between the P and Q results of two scenarios is measured by the fitting degree (FD), as follows,

$$FD = 1 - \frac{\sum(y_2 - y_1)^2}{\sum(y_1 - \text{mean}(y_1))^2} \quad (9)$$

where $y_i$ is the time series of P or Q in Scenario $i$, $i$=1, 2, and mean($y_1$) is the mean value of $y_1$. The average of the FD of P and that of Q is regarded as the FD of one validation case.

The validation results of the example identification case in the Fault YC event are given in Fig. 7. The FD of this case is 1, which means the dynamics of the identified load models are very similar to the actual load models. In addition, it can be observed that the results of both the induction motor part and the static load part in Fig. 7 are very similar, which means the identified static load model has similar dynamics to the actual model, even though the parameters are not the same.

*2) Examples with the Impact of Measurement Errors*
  *a) Identification*

In this section, the PMU measurement errors are considered. The measurement errors here are based on the experimental test results in [16], which include two parts, i.e., the systematical error (the offset) and the accidental error (the random variations). The signal to noise ratio (SNR) is used to describe the relative magnitudes of signal and noise, as follows,

$$SNR = \frac{Energy\_of\_signal}{Energy\_of\_noise} = 10\log\frac{\sum y_i^2}{\sum e_i^2} \quad (10)$$

where $y_i$ is the signal without error and $e_i$ is the measurement error. In this paper, the average of the SNR of P and that of Q is used to evaluate the SNR of one LMC. To better analyze the impact of measurement errors, more SCs are conducted under three different disturbance levels (DLs), which are denoted as DL=1, 2, 3. The ambient signals with different DLs are generated by adjusting the magnitudes of the low-pass filtered white noise which is penetrated at the load buses. With a larger DL, the measurement data has larger SNR. The 50 SCs in the previous subsection are DL=1 cases, and 50 DL=2 cases and 50 DL=3 cases are also simulated in the same way.

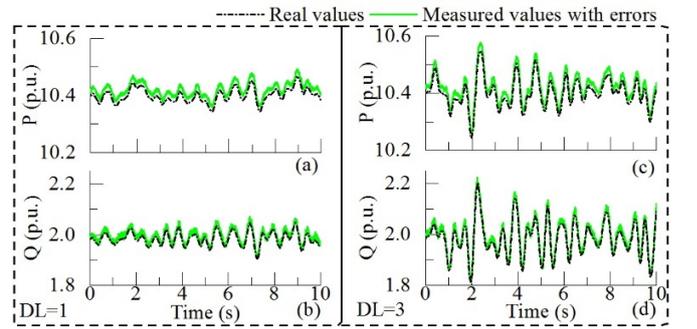

Fig. 8 Impact of measurement errors on P and Q measurement data

Two examples with the impact of errors are given. Example 1 is the measurement data of Bus HZ in SC 3 of DL=1 cases, while Example 2 is the measurement data of Bus HZ in SC 3 of DL=3 cases. The real values and the measured values with errors of P and Q in two examples are given in Fig. 8. The SNR of two examples are 8.20 and 25.24, respectively. It can be observed that the ambient signals with a larger DL have relatively larger disturbance magnitudes and larger SNR.

A low-pass filter with the cut-off frequency as 2Hz is used to pre-process the measurement data, after which load model parameters are identified from the filtered data. In Example 1, $D_{real}$=[43.98 17.89 0.63 3.40] and $D_{opt}$=[44.84 18.57 0.60 3.38]. In Example 2, $D_{real}$=[58.97 21.40 0.79 5.05] and $D_{opt}$=[59.75 21.44 0.82 4.99]. To conclude, the identification results deviate a little from the actual parameters, but they are still very close.

*b) Validation*

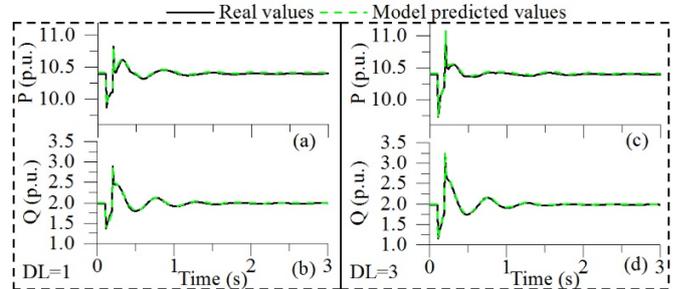

Fig. 9 Validation resutls of Bus HZ: a DL=1 case and a DL=3 case

The identified load models in the previous subsection are also validated in Fault YC event, the results of which are given in Fig. 9. The FDs of the DL=1 case and the DL=3 case are 0.978 and 0.983, respectively. The results have shown that the load models identified from the data with errors still have similar dynamics to the actual models under large disturbance events. It can also be concluded that different load models will lead to different response results even at the same OP and in the same event, which validates the necessity of load modeling.

## C. Results of Quasi-convexity and SQP Reliability Tests

### 1) Quasi-convexity Test

In this subsection the quasi-convexity of the OF is tested. 1000 DL=1 LMCs are used in the test. For one LMC, 10,000 pairs of points ($D_{2*i-1}$, $D_{2*i}$), $i$=1, 2, …, 10,000 are randomly selected in the feasible region of the decision variables. If OF(($D_{2*i-1}$+$D_{2*i}$)/2)<max{OF($D_{2*i-1}$), OF($D_{2*i}$)} holds, it is regarded as "successful". In this way, the success percentage (SP) of all the pairs of points for this LMC is obtained.

All the 1000 LMCs are tested in this way, after which the histogram of 1000 SPs is given in Fig. 10. In this histogram, the range of the SPs is [88, 100], and it is divided by the step of 0.1. The SPs of 953 LMCs are larger than 96%. The median SP is 98.69%, and the minimal SP is 88.28%. To conclude, the quasi-convexity test is successful for most of the pairs of points, which can provide the basis for the application of the gradient-based optimization algorithms, such as the SQP.

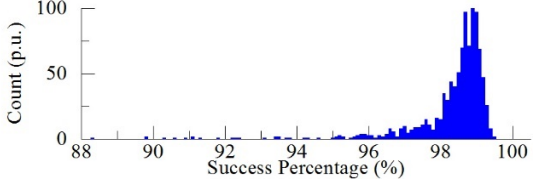

Fig. 10 Histogram of success percentage of quasi-convexity test results.

### 2) SQP Reliability Test

The reliability of the SQP to achieve the GOSs is tested. For one LMC, 1000 times of SQP optimization from 1000 different randomly generated IFSs are repeated, the solutions of which are denoted as $D_i$, $i$=1, 2, …, 1000. Afterwards, the GOS $D_{opt}$ is regarded as argmin(OF($D_i$)). The $i$th SQP case is regarded as successful if $\sum_{j=1}^{4}(D_{i,j}/D_{opt,j} - 1) < 0.1$. Then, the SP of all the 1000 times of SQP for this LMC can be obtained. Among the SPs of all the 1000 LMCs, 961 of them are 100%, and 35 of them are 99.9%, which means only 4 SPs are less than 99.9%. The minimal SP among them is 95.7%. The results have shown the strong ability of the SQP algorithm to achieve the GOSs.

To further improve the reliability, one approach is to repeat the SQP from different randomly generated IFSs and select the best solution as the final one. In the following sections, the results are all obtained through 3 times of repeated SQP.

### 3) An Example of Objective Function

The OF of Example 1 in Section VI.B.2.a is given here. With $D_1$=[26.34 10.89 1.45 5.99] and $D_2$=[67.32 29.52 1.95 6.03], a two dimensional surface is generated as: $D(k_1, k_2)$=$D_{opt}$+$k_1$($D_1$-$D_{opt}$)+$k_2$($D_2$-$D_{opt}$). Then, the contour map of $\log_{10}$(OF($D(k_1, k_2)$)) is given in Fig. 11. The values of the points which are outside the feasible region are set as 6 in this figure. It can be observed that the OF is quasi-convex in most part of the feasible region, which is consistent with the quasi-convexity test results. However, for some part which is far from $D_{opt}$ or around the boundary of the feasible region, the OF is not quasi-convex.

An example of a pair of points which has failed the quasi-convexity test is also given, which is $D_3$(0.62, 0.60) and $D_4$ (1.33, -0.06). With OF($D_3$)=OF($D_4$), OF(($D_3$+$D_4$)/2) is larger than them because it is outside the contour line. However, with $D_3$, $D_4$ and ($D_3$+$D_4$)/2 being the IFSs, the solutions by the SQP are all $D_{opt}$. This is an example that the GOS can still be achieved even the IFS is located in the region where the OF is not quasi-convex. This can explain the phenomenon that most of the SPs in the SQP reliability tests are larger than 99.9%, while the SPs of the quasi-convexity tests are less than 100%. For this case, the SP of the quasi-convexity test is 98.82%, while that of the SQP reliability test is 100%. To conclude, the quasi-convexity of the OF can help to explain why the SQP can achieve the GOS, but it is not a necessary condition that the OF must be quasi-convex in the whole feasible region.

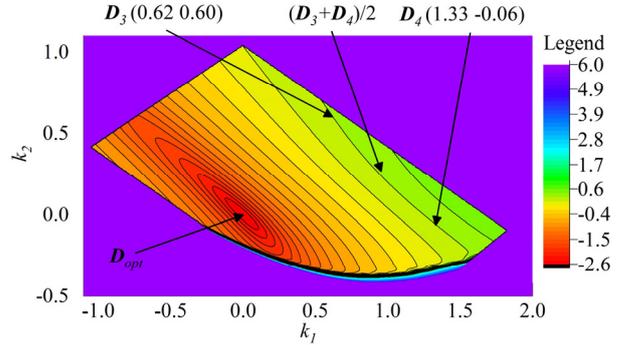

Fig. 11 An example of objective function

## D. Identification and Validation Results

### 1) Identification Accuracy

In this section, all the 3000 LMCs with 3 different DLs are conducted with the impact of measurement errors. After obtaining the $D_{opt}$ of one LMC, each of the parameters in $D_{opt}$ is firstly divided by the same parameter in $D_{real}$ for normalization, which means 1 is the accurate value for one parameter. Then, the mean SNRs, the mean values ($\mu$) and the standard deviations ($\sigma$) of each parameter of all the 3000 LMCs with 3 different DLs are calculated, which are given in TABLE I.

The identification results are accurate even with the impact of measurement errors for all the DLs, because the $\mu$s are close to 1 and the $\sigma$s are much smaller. With the increase of DL, the $\mu$s are closer to 1 and the $\sigma$s are smaller, which means more accurate identification results are obtained with larger DL. The results have validated that the proposed approach has the robustness towards the impact of measurement errors.

TABLE I Mean Values and Standard Deviations of the Identification Results

| DL | Mean SNR | $\sigma(a)$ | $\mu(a)$ | $\sigma(b)$ | $\mu(b)$ | $\sigma(H_2)$ | $\mu(H_2)$ | $\sigma(T_m)$ | $\mu(T_m)$ |
|---|---|---|---|---|---|---|---|---|---|
| 1 | 14.24 | 0.030 | 1.001 | 0.029 | 1.005 | 0.034 | 0.993 | 0.021 | 1.000 |
| 2 | 21.77 | 0.022 | 0.999 | 0.024 | 1.002 | 0.028 | 0.999 | 0.016 | 1.000 |
| 3 | 28.03 | 0.020 | 0.999 | 0.022 | 1.001 | 0.023 | 1.000 | 0.014 | 1.000 |

DL: disturbance level. SNR: signal to noise ratio.

### 2) Validation Accuracy under Large Disturbance Events

All the 150 SCs of 3 different DLs are validated under the two fault events according to the process in Section VI.B.1.b. Afterwards, 20 FDs for 20 load models are calculated for one SC, after which 1000 FDs are calculated for one DL. The median FDs, the percentages of the FDs which are larger than 0.95 and 0.9, are given in TABLE II.

It can be observed from the results that the median FDs are close to 1, and the percentages are large. In addition, the results tend to be more accurate with the increase of DL. It is validated here that the identified load models have performed similarly to the actual load models under large disturbance events.

TABLE II Fitting Degree Results of the Validation Cases

| DL | Fault YC | | | Fault XLS | | |
|---|---|---|---|---|---|---|
| | M(FD) | %(FD>0.9) | %(FD>0.95) | M(FD) | %(FD>0.9) | %(FD>0.95) |
| 1 | 0.981 | 85.5 | 76.4 | 0.977 | 89.7 | 79.4 |
| 2 | 0.988 | 97.5 | 94.2 | 0.984 | 98.5 | 95.8 |
| 3 | 0.991 | 99.0 | 97.4 | 0.987 | 99.1 | 97.9 |

DL: disturbance level. FD: fitting degree. M(FD): median value of FD.

*E. Comparison with Previous ASLM Approach*

TABLE III Comparison with previous ASLM approach

| Approach | Quasi-convexity | Optimization algorithm | Computation time | Model structure |
|---|---|---|---|---|
| Previous | No | Heuristic | >30 s | Z+IM |
| Proposed | Mostly | Gradient based | 1.82 s | ZIP/EX+IM |

The improvements of the proposed approach compared with the previous ASLM approach in [11] are summarized in TABLE III and discussed as follows.

Firstly, through the transformation of the induction motor model, the OF is mostly quasi-convex in the feasible region. In comparison, an example of the OF in the previous approach is given in Fig. 12, which is based on the results in [11]. It is clear that the OF here is non-convex. Based on the quasi-convexification of the OF, gradient-based algorithms can be applied in solving the optimization problem, which can significantly improve the computation efficiency. With the data length being 10 s, The average identification time for one LMC of all the 1000 DL=1 LMCs is 1.82 s, while it takes more than 30 s by the previous approach according to the results in [11]. Since the computation time is much shorter than the data length, real-time continuous periodical ASLM, in which the data length is no shorter than the identification period, can be achieved.

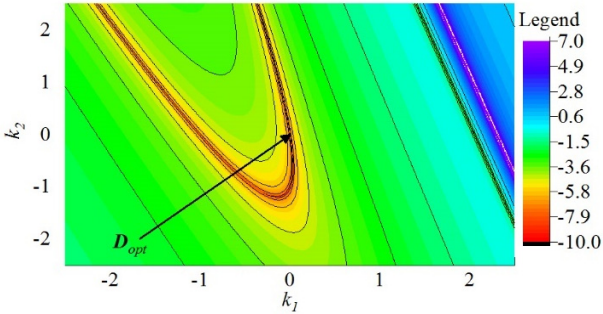

Fig. 12 An example of objective function in previous ASLM approach

Secondly, based on the proposed hierarchical identification framework, more complicated static load model structures, such as the widely used ZIP model, can be applied in ASLM. In the previous approach, due to the limited computation ability of heuristic optimization, only the constant Z model can be applied. This is because the parameters of the static load model are identified through regression in the proposed hierarchical framework. In this way, the number of parameters which are identified through optimization is significantly reduced, which also contributes to the improvement of computation efficiency.

## VII. CONCLUSION

In this paper, a hierarchical framework to identify load model parameters from ambient signals measurements is proposed. The static load parameters are identified through regression in the lower stage, and the induction motor parameters are identified through optimization in the upper stage with the regression residuals as the OF. The case study results have shown that the proposed algorithm can accurately identify the load models with the existence of measurement errors, and the identified load models have similar dynamic performance with the actual load models under large disturbance events. The quasi-convexity of the OF and the reliability of the SQP are also validated through case study. This is the first time that the hidden quasi-convexity of load modeling is explored, which can provide the basis for the application of gradient-based algorithms. Compared with the previous ASLM approaches, the model structure complexity and computation efficiency are both improved. In this way, the practical applicability of ASLM is significantly enhanced.